\documentclass[%
 reprint,
 amsmath,amssymb,
 aps,
]{revtex4-2}

\usepackage{graphicx}
\usepackage{dcolumn}
\usepackage{bm}
\usepackage{float}
\usepackage{ragged2e} 

\begin{document}

\preprint{APS/123-QED}

\title{Non-resonant ultra-fast multipactor regime in dielectric-assist accelerating structures}

\author{Daniel González-Iglesias}
 \email{daniel.gonzalez-iglesias@uv.es}
\author{Benito Gimeno}
\author{Daniel Esperante}
 \altaffiliation[Also at ]{Electronics Engineering
Department, Universitat de València, 46100 Burjassot, Spain}
\author{Pablo Martinez-Reviriego}
\author{Pablo~Martín-Luna}
\author{Nuria Fuster-Martínez}
\author{César Blanch}
\author{Eduardo Martínez}
\author{Abraham Menendez}
\author{Juan Fuster}
\affiliation{Instituto de Física Corpuscular (IFIC), CSIC-UV, c/ Catedrático José Beltrán 2, 46980 Paterna, Spain}%

\author{Alexej Grudiev}

\affiliation{CERN, European Organization for Nuclear Research, Geneva 1221, Switzerland}

\date{\today}

\begin{abstract}
The objective of this work is the evaluation of the risk of suffering a multipactor discharge in an S-band dielectric-assist accelerating (DAA) structure for a compact low-energy linear particle accelerator dedicated to hadrontherapy treatments. A DAA structure consists of ultra-low loss dielectric ciylinders and disks with irises which are periodically arranged in a metallic enclosure,  with the advantage of having an extremely high quality factor and very high shunt impedance at room temperature, and it is therefore proposed as a potential alternative to conventional disk-loaded copper structures. However, it has been observed that these structures suffer from multipactor discharges. In fact, multipactor is one of the main problems of these devices, as it limits the maximum accelerating gradient. Because of this, the analysis of multipactor risk in the early design steps of DAA cavities is crucial to ensure the correct performance of the device after fabrication. In this paper, we present a comprehensive and detailed study of multipactor in our DAA design through numerical simulations performed with an in-house developed code based on the Monte-Carlo method. The phenomenology of the multipactor (resonant electron trajectories, electron flight time between impacts, etc.) is described in detail for different values of the accelerating gradient. It has been found that in these structures an ultra-fast non-resonant multipactor appears, which is different from the types of multipactor theoretically studied in the scientific literature. In addition, the effect of several low electron emission coatings on the multipactor threshold is investigated. Furthermore, a novel design based on the modification of the DAA cell geometry for multipactor mitigation is introduced, which shows a significant increase in the accelerating gradient handling capabilities of our prototype.

\end{abstract}

\maketitle


\section{\label{introduction} Introduction}

Multipactor breakdown is an undesirable discharge phenomenon that occurs in devices operating under high vacuum conditions and high-power radiofrequency (RF) electromagnetic fields \cite{Vaughan}. When certain resonant conditions are met, free electrons within the device are accelerated by the RF electric field and collide with the metallic walls. If the kinetic energy at impact is sufficient, secondary electrons may be released from the surface, triggering a chain reaction that exponentially increases the electron population inside the device.

The onset of multipactor discharge leads to various detrimental effects, including increased signal noise and reflected power, heating of device walls, outgassing, detuning of resonant cavities, and surface damage. Thus, this phenomenon poses a significant limitation on the maximum RF power handling capability of devices. 

To address the challenges posed by multipactor discharge, researchers focus on designing materials and geometries that minimize its impact. Additionally, efforts are underway to develop techniques for suppressing or eliminating multipactor altogether. Understanding and mitigating this phenomenon are crucial for optimizing device performance and ensuring reliable operation in high-power RF applications.

Multipactor can occur in a wide range of scenarios, such as passive components in satellite communication payloads, klystrons and particle accelerators \cite{Shemelin}.

Recently, dielectric loaded accelerating structures are being proposed as a potential alternative to
conventional disk-loaded copper structures. These dielectric assist accelerating (DAA) structures consists of ultra-low lossy dielectric ciylinders and disks with irises which are periodically arranged in a metallic enclosure \cite{Satoh}. The advantage of the DAA structure is that it has an extremely high quality factor and a very high shunt impedance at room temperature since the electromagnetic field distribution of the accelerating mode $TM_{02}$ reduces greatly the losses on the metallic wall as compared to that of a $TM_{01}$ mode of the pillbox cylindrical cavity. Several DAA designs have been presented in the literature, and even some prototypes have been build and tested \cite{Satoh2}, \cite{Mori}, \cite{9387344}. Although their good performance in terms of quality factor and shunt impedance has been demonstrated with measurements, it has also been found that DAA structures suffer multipactor during operation. So far, the maximum accelerating field of DAA cavities has been limited to a few MV/m by discharges. In \cite{Mori}, it is stated that in the DAA cavity, with the nominal Secondary Electron Yield (SEY) of the dielectric, the multipactor limits the maximum accelerating gradient to about $1$ or $2$ MV/m. If a low-SEY coating is applied to the dielectric, they found that the gradient can be increased up to $12$ MV/m. However, they also state that for the application of the DAA cavity to the accelerator facility, a higher accelerating gradient would be required, e.g., $30$ MV/m. To the Authors' knowledge, the risk of multipactor occurrence in the DAA structure described in \cite{Mori} was not taken into account during its design. Then, after fabrication of the prototype it was found that the multipactor was a major concern for good performance and, even with the application of the low-SEY coating, the maximum gradient was still limited by discharges. 

Therefore, the importance of risk assessment of the multipactor effect in the design phases of DAA structures is demonstrated. In this work, we present a comprehensive analysis of the multipactor phenomenon in the design of an S-band DAA for a compact low-energy linear particle accelerator dedicated to hadrontherapy treatments \cite{Pablo_DAA}. This device is being developed in collaboration between IFIC and CERN. To perform the multipactor studies, an in-house code based on the Monte-Carlo algorithm has been developed. The program tracks the trajectories of the electrons inside the DAA cell and takes into account the emission of secondary electrons after a primary electron hits the surfaces of the device. Thus, the population of electrons over time is recorded and, with this information, the presence or absence of the multipactor discharge is determined.  

This document is organized as follows. Section \ref{DAA_description} describes the DAA structure for which the multipactor analysis is performed. Next, Section \ref{multipactor_code} explains the characteristics of the multipactor simulation code. Next, the results of the multipactor simulations are detailed in Section \ref{multipactor_simulations}. For a better understanding of the results of the simulations, a simple theoretical model based on analytical expressions, which can be obtained from solving the differential equations of motion of the electron, is shown in Section \ref{an_model}. The phenomenology of the multipactor in the cavity (types of electron resonant trajectories, zones of the cell where the discharge appears, time of flight between successive impacts of the electron with the walls, angle of impact with respect to the surface) is discussed in Section \ref{phenomenology} with the help of the theoretical model of Section \ref{an_model}. This will provide insight into the characteristics of the novel non-resonant ultra-fast multipactor found in these structures, which differs from classical multipactor theory in which the time of flight of electrons between successive impacts with the device walls is required to be an odd (even) number of half-periods of the RF signal for double (single) surface multipactor regimes.  As it will be seen later, multipactor discharge limits the performance of our DAA design. Because of this, in Section \ref{mod_cell} a novel design based on the modification of the DAA cell geometry for multipactor mitigation is introduced, showing a significant increase in the accelerating gradient handling capabilities of our prototype. Finally, the main conclusions of this study are presented in Section \ref{conclusions}.

\section{\label{DAA_description} Description of the DAA structure}

\begin{figure}[H]
\centering
\includegraphics[scale=0.30]{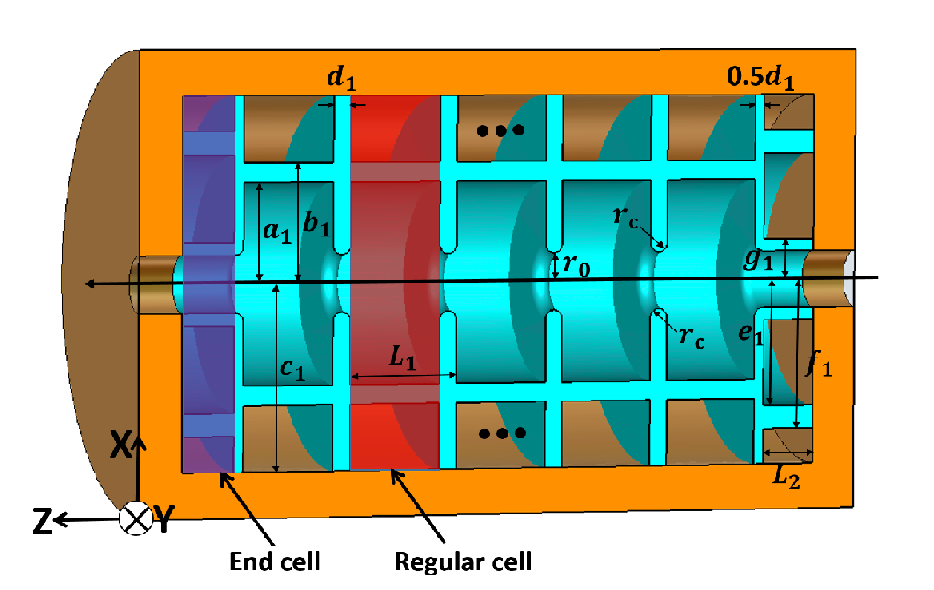}
\caption{Schematic of a DAA structure with its characteristic dimensions. Figure extracted from \cite{9387344}.  \label{DAA_scheme}}
\end{figure}

The DAA cell in which we are going to study the multipactor is a design for a linear accelerator of low-energy particles (such as protons or carbon ions) for handrontherapy treatments \cite{Pablo_DAA}. This device has been conceived to operate in standing wave with the mode of operation $\pi$-$TM_{02}$ at a frequency of $f= 2998.1$ MHz (S-band), for an acceleration coefficient $\beta = 0.4$, $\beta$ being the axial velocity normalized with the speed of light in vacuum $c$. The schematic of the DAA structure is shown in the Fig. \ref{DAA_scheme}. 

\begin{figure}[H]
\centering
\includegraphics[scale=0.25]{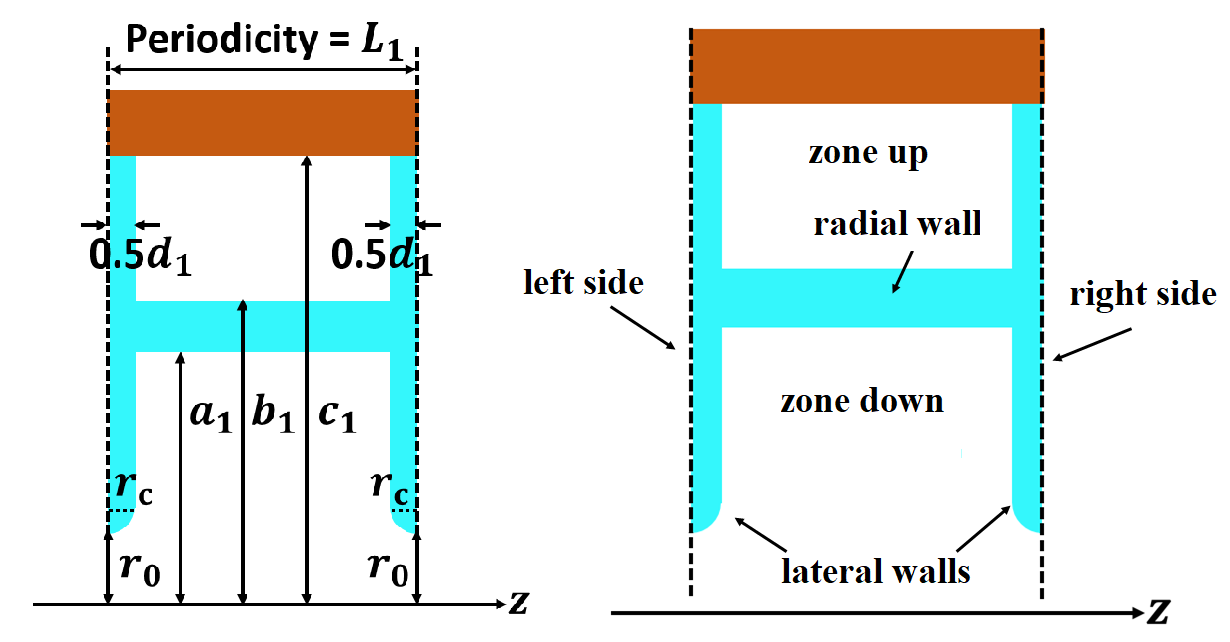}
\caption{Left: schematic of a regular cell of a
DAA structure. Right: nomenclature with the different zones of the DAA cell \cite{9387344}.  \label{cell_zonas}}
\end{figure}

The structure has rotational symmetry around the z-axis. We will focus our multipactor study on a single regular cell. The outer wall of the cell is metallic (copper) with a radius $c_1$, the inner and outer radii of the dielectric are $a_1$ and $b_1$, respectively, the iris radius is $r_0$ and its radius of curvature $r_c$, $d_1$ is the thickness of the dielectric disc, the length of the cell is $L$, which we assume to be periodic. The dimensions of the cell are shown in the Table \ref{dimensiones}. In the Fig. \ref{cell_zonas} is shown the schematic of the regular cell and it is defined the nomenclature that we will use to name the
different parts of the cell when analysing the multipactor, which has been divided in zone down and zone up.

\begin{table}[H] \caption{Dimensions of the regular cell of the DAA structure}
\begin{center}\label{dimensiones} 
\begin{tabular}{|c|c|}
\hline
Dimension & Value (cm) \\ \hline
       $r_0$   &  0.2          \\ \hline
    $r_c$      &  0.4284          \\ \hline
    $a_1$     &  4.8466          \\ \hline
    $b_1$      &  5.5755          \\ \hline
    $c_1$      &  8.7439          \\ \hline
    $d_1$      &  0.8569          \\ \hline
       $ L_1$      &  1.9986          \\ \hline
\end{tabular}
\end{center}
\end{table}


The dielectric of the cell is $MgTiO_3$ and has a relative dielectric permittivity of
$\epsilon_r =16.66$; we will assume that it has no ohmic losses. For the Secondary Electron Yield \cite{Reimer} (SEY) coefficient of this material there are experimental measurements which are described in \cite{Report_SEY}. In addition, several coatings that can be made on the dielectric in order to reduce the emission of
secondary electrons are detailed in this reference. Four cases will be considered in the
multipactor simulations presented in this work: dielectric without coating (sample 7 in \cite{Report_SEY}), dielectric with
SEY-reducing coating applied by Acree technologies inc. \cite{us_company} (sample 8), dielectric with coating
applied by Nanotec Co. \cite{japanese_company} (sample 9), and dielectric with ac-400 nm coating applied by
CERN (bldg 867, sample 2). It should be noted that samples 8 and 9 correspond to the same
type of coating, although the aim is to study the possible differences in the final SEY of the
material depending on the particular procedure applied by each institution.

\begin{figure*}
\centering
\includegraphics[scale=0.4]{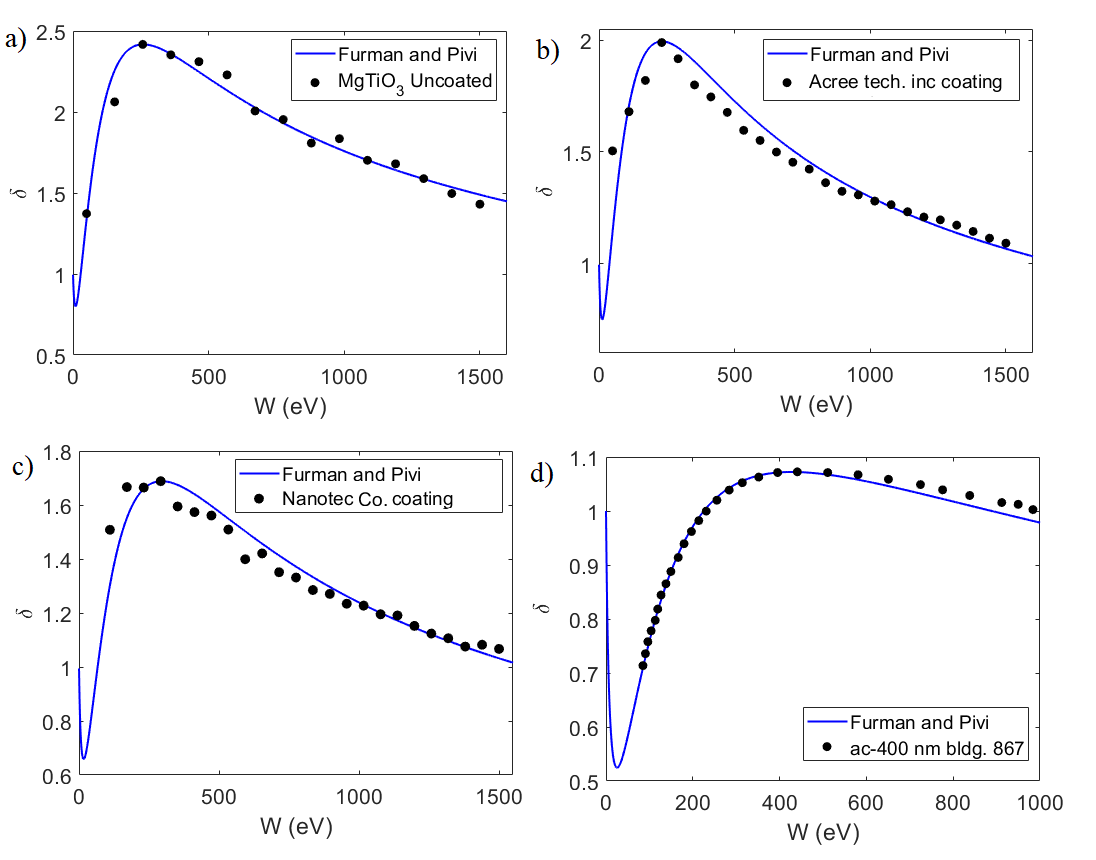}
\caption{SEY of $MgTiO_3$ samples without coating (sample 7) (a), $MgTiO_3$ with Acree technologies inc. coating (b), with Nanotec Co. coating (c), and with ac-400nm bldg 867 coating (d), as a function of the primary electron kinetic energy. Experimental data as well as the fit to the Furman and Pivi model are shown.  \label{seys}}
\end{figure*}

For each of the materials under consideration, measurements of the SEY curve at normal incidence were made at different physical positions of the sample in order to determine possible variations of the data from one point to another. For example, in the case of the uncoated dielectric we have measurements for three positions. To obtain the experimental SEY curve we simply proceed by calculating the average value of the different SEY curves for each incident electron energy. Then we will fit this experimental curve to the theoretical model of Furman and Pivi \cite{Furman}, \cite{Lara} that will allow us to obtain the different parameters of the SEY model that we will implement in our simulator. Fig. \ref{seys} shows the experimental data and fits for the dielectric samples without coating (a), with Acree technologies inc. coating (b), with Nanotec Co. coating (c), and with ac-400 nm bldg 687 coating (d).

The main parameters of the SEY for each dielectric material are shown in the Table \ref{parametros_SEY}. The parameter $W_1$ represents the electron impact kinetic energy at normal incidence above which the SEY is greater than unity, $W_{max}$ represents the impact kinetic energy (at normal incidence) for which the SEY coefficient takes its maximum value, which is given by $\delta_{max}$.

\begin{table}[H] \caption{Main SEY parameters}
\begin{center}\label{parametros_SEY} 
\begin{tabular}{|c|c|c|c|}
\hline
Material & $W_1$ (eV) & $W_{max}$ (eV) & $\delta_{max}$  \\ \hline
$MgTiO_3$ without coating        & 28.6  & 257.1 & 2.41 \\ \hline
$MgTiO_3$ Acree tech. inc. coating  & 38.6 & 231.2 & 1.98 \\ \hline
 $MgTiO_3$ Nanotec Co. coating   & 66.6 & 291.7 & 1.69 \\ \hline
  $MgTiO_3$ ac-400 nm coating   & 225.5 & 428.0 & 1.07 \\ \hline
\end{tabular}
\end{center}
\end{table}

For the metallic copper wall in the \textit{up} region we will take the SEY parameters of \cite{RFgun}, with $\delta_{max} = 2.44$, $W_1 = 13$ eV, $W_{max}= 200$ eV.

\section{\label{multipactor_code} Multipactor simulation code}

The multipactor risk assessment is carried out by Monte-Carlo numerical simulations of the trajectories of an electron cloud in the device cavity, appropriately modelling the secondary emission of electrons from the walls of the component as a consequence of the impact of a primary electron. For this purpose, a multipactor effect
simulation code adapted for this type of structures has been developed in-house. This code is based on the single electron model, in which each particle tracked during the simulation represents a single ``real-world''
electron. The RF electromagnetic field in the DAA structure is calculated using the Superfish software \cite{SUPERFISH}, which provides the electric and magnetic field corresponding to the mode of operation in a 2-D mesh in $r-z$ coordinates (no dependence on azimuthal angle) that is imported by the multipactor simulation code and, by means of bilinear interpolation, allows to compute the Lorentz force on the positions occupied by the electrons inside the component. By equating this force to the relativistic version of Newton's Second Law, the 3-D trajectories of each particle can be solved numerically using the Higuera-Cary algorithm \cite{Ripperda_2018}.  In this code the space-charge
effect of the electron cloud is neglected since it is of minor importance in the first steps of the growth of the multipactor discharge \cite{Semenov_modulaciones}. After the update of the particle position at each time integration step, it is checked whether the particle collides with any wall of the structure. When an electron collides with the component walls, there are four possible consequences of the interaction: elastic reflection, inelastic reflection, emission of true secondaries and absorption. Furman and Pivi's Secondary Electron Yield model described in \cite{Lara} is used to characterise the probabilities of each of the above interaction types. In elastic and inelastic reflection, the electron is re-emitted back into the cavity at the same angle as at the instant of impact. In the case of elastic reflection, the kinetic energy of the electron at the entrance and at the release are equal, while for inelastic reflection a certain amount of energy is lost. Regarding the true secondary electron emission type interaction, one or more secondary electrons are emitted. Each of these secondary electrons departs from the surface with an angle following the probability distribution given by the cosine law \cite{coseno}, while the energy follows the distribution functions detailed in \cite{Lara}, which take into account the principle of energy conservation (i.e., the sum of the total energy of the emitted electrons is equal or less than the energy of the impacting electron). If the electron is absorbed it is removed from the simulation. Once a predefined number of periods of the RF signal have been analyzed, the simulation concludes and the population of electrons within the device is plotted over time. This information is crucial in determining whether or not a multipactor discharge is likely to occur. Concretely, if a sustained growth in the electron population is observed over time, then the multipactor discharge will appear.

\section{\label{multipactor_simulations} Multipactor simulations}

We have performed multipactor simulations for the DAA elemental cell with the uncoated dielectric and for the three low-SEY coatings described above. The simulations explore a wide range of RF electric field amplitudes (or accelerating gradients) $E_0$ in the range $0.01-200$ MV/m, being $E_0$ the maximum amplitude of the RF electric field in the axis of the cell, which is reached in the central point. The DAA cell has been divided into two zones for the multipactor simulations. This division is motivated by the fact that the electrons cannot move from one region to another and, therefore, a better understanding of the specific zones of the device where the discharge occurs can be obtained by separate multipactor simulations in each zone.  Nevertheless, the amplitude of the RF electromagnetic field in the cavity can be described by the amplitude $E_0$ in both zones.

First, we present the results of the multipactor simulations for the DAA structure with the uncoated dielectric in Table \ref{tabla_multipactor}. In addition to results from our in-house code, we also include CST Particle Studio \cite{CST} simulations in order to check the correct operation of our algorithm. As it can be seen, both codes predict the presence of the discharge in the same cases with excellent agreement. Therefore, the verification of our multipactor algorithm is successfully demonstrated. According to the simulations, the presence of multipactor in the DAA structure is verified for amplitudes $E_0$ between $1$ and $200$ MV/m. The multipactor appears first in the \textit{down} zone for an electric field amplitude of $E_0=1$ MV/m. In the \textit{up} zone, the discharge is expected for fields above $E_0>5$ MV/m.

\begin{table}[!h] \caption{Results of multipactor simulations for the dielectric $MgTiO_3$ without coating (\checkmark indicates multipactor, $\times$ means no discharge).}
\begin{center}\label{tabla_multipactor} 
\begin{tabular}{|c|cc|cc|}
\hline
\multicolumn{1}{|l|}{} & \multicolumn{2}{c|}{Up}               & \multicolumn{2}{c|}{Down}               \\ \hline
$E_0$ (MV/m)            & \multicolumn{1}{c|}{In-house code} & CST & \multicolumn{1}{c|}{In-house code } & CST \\ \hline
0.01                    & \multicolumn{1}{c|}{$\times$}                  & $\times$     & \multicolumn{1}{c|}{$\times$}                  & $\times$     \\ \hline
0.1                  & \multicolumn{1}{c|}{$\times$}                    &  $\times$       & \multicolumn{1}{c|}{$\times$}                    &   $\times$      \\ \hline
1                   & \multicolumn{1}{c|}{$\times$}                    &   $\times$      & \multicolumn{1}{c|}{$\checkmark$}                    &  $\checkmark$       \\ \hline
5                    & \multicolumn{1}{c|}{\checkmark}                    &     \checkmark  & \multicolumn{1}{c|}{\checkmark}                    &       \checkmark  \\ \hline
10                    & \multicolumn{1}{c|}{\checkmark}                    &    \checkmark     & \multicolumn{1}{c|}{\checkmark}                    &        \checkmark \\ \hline
20                    & \multicolumn{1}{c|}{\checkmark}                    &     \checkmark    & \multicolumn{1}{c|}{\checkmark}                    &       \checkmark  \\ \hline
50                     & \multicolumn{1}{c|}{\checkmark}                    &  \checkmark     & \multicolumn{1}{c|}{\checkmark}                    &  \checkmark       \\ \hline
75                     & \multicolumn{1}{c|}{\checkmark}                    &   \checkmark     & \multicolumn{1}{c|}{\checkmark}                    &       \checkmark  \\ \hline
100                      & \multicolumn{1}{c|}{\checkmark}                    &  \checkmark      & \multicolumn{1}{c|}{\checkmark}                    &       \checkmark  \\ \hline
125                   & \multicolumn{1}{c|}{\checkmark}                    &   \checkmark      & \multicolumn{1}{c|}{\checkmark}                    &       \checkmark   \\ \hline
150                   & \multicolumn{1}{c|}{\checkmark}                    &   \checkmark      & \multicolumn{1}{c|}{\checkmark}                    &       \checkmark \\ \hline
175                  & \multicolumn{1}{c|}{\checkmark}                    &   \checkmark      & \multicolumn{1}{c|}{\checkmark}                    &       \checkmark   \\ \hline
200                   & \multicolumn{1}{c|}{\checkmark}                    &   \checkmark      & \multicolumn{1}{c|}{\checkmark}                    &       \checkmark   \\ \hline
\end{tabular}
\end{center}
\end{table} 

The results of the multipactor simulations for the dielectric with Nanotec Co. coating and with Acree technologies inc. coating are identical to those of the uncoated dielectric shown above. Thus, the multipactor thresholds are $1$ MV/m in the \textit{down} zone and $5$ MV/m in the \textit{up} zone. Therefore, neither the Nanotec Co. coating nor the Acree technologies inc. coating are able to inhibit the multipactor discharge for any of the electric field amplitudes studied.

The multipactor simulations results for the dielectric with the ac-400nm bldg 867 coating are shown in Table \ref{tabla_multipactor_bldg687}. The obtained results differ from those of the uncoated dielectric, showing ranges of electric field amplitudes for which the coating is able to inhibit the multipactor discharge. This effect appears mainly in the \textit{down} region. For this region and with the coating, the multipactor threshold appears at $E_0 = 20$ MV/m, while without the coating the threshold was only $1$ MV/m, and the multipactor appearance was continuous for any value of $E_0$ up to the upper limit of $200$ MV/m. The application of the coating allows multipactor-free windows to be opened in the \textit{down} zone in the ranges $35-45$ MV/m, $65-125$ MV/m, and $145-165$ MV/m. This is therefore a considerable improvement over the uncoated case. However, these beneficial effects of the coating are not as significant in the \textit{up} zone, where the coating only raises the threshold to $10$ MV/m compared to $5$ MV/m in the uncoated case, and is not able to generate additional multipactor-free windows, unlike what is found in the \textit{down} zone. 

\begin{table}[H] \caption{Results of multipactor simulations for the D16 dielectric $MgTiO_3$ with ac-400 nm bldg 867 coating (\checkmark indicates multipactor, $\times$ means no discharge)}
\begin{center} \label{tabla_multipactor_bldg687} 
		\begin{tabular}{|c|c|c|} 
			\hline
			$E_0$ (MV/m)            & Up   & Down  \\ \hline
			0.01                    & $\times$  & $\times$     \\ \hline
			0.1                    & $\times$  & $\times$   \\ \hline
			1                    & $\times$  & $\times$   \\ \hline
			5                   & $\times$   & $\times$  \\ \hline
            10                   & \checkmark   & $\times$  \\ \hline
   		15                    & \checkmark   &  $\times$   \\ \hline
        	20                    & \checkmark   &  \checkmark   \\ \hline
            25                    & \checkmark   &  \checkmark    \\ \hline
             30                    & \checkmark   &  \checkmark   \\ \hline
            35                    & \checkmark   &  $\times$   \\ \hline
             40                    & \checkmark   &  $\times$   \\ \hline
             45                    & \checkmark  &  $\times$   \\ \hline
             50                    & \checkmark   &  \checkmark    \\ \hline
             55                    & \checkmark   &  \checkmark    \\ \hline
             60                    & \checkmark   &  \checkmark   \\ \hline
             65                    & \checkmark   &  $\times$   \\ \hline
             70                    & \checkmark   &  $\times$   \\ \hline
             75                    & \checkmark  &  $\times$   \\ \hline
             80                    & \checkmark   &  $\times$   \\ \hline
             85                    & \checkmark   &  $\times$   \\ \hline
             90                    & \checkmark   &  $\times$   \\ \hline
             95                    & \checkmark  &  $\times$   \\ \hline

		\end{tabular}
  \quad \quad 
  		\begin{tabular}{|c|c|c|} 
			\hline
			$E_0$ (MV/m)            & Up   & Down  \\ \hline
             &    &     \\ \hline
             100                    & \checkmark   &  $\times$   \\ \hline
            105                   & \checkmark   & $\times$     \\ \hline
			110                    & \checkmark   & $\times$   \\ \hline
			115                  & \checkmark  & $\times$   \\ \hline
			120                   & \checkmark   & $\times$   \\ \hline
   		125                    & \checkmark  &  $\times$   \\ \hline
        	130                    & \checkmark   &  \checkmark    \\ \hline
            135                    & \checkmark   &  \checkmark    \\ \hline
             140                    & \checkmark   &  \checkmark    \\ \hline
            145                    & \checkmark  &  $\times$   \\ \hline
             150                    & \checkmark   &  $\times$   \\ \hline
             155                    & \checkmark  &  $\times$   \\ \hline
             160                    & \checkmark  &  $\times$   \\ \hline
             165                    & \checkmark   &  $\times$   \\ \hline
             170                    & \checkmark  &  \checkmark    \\ \hline
             175                    & \checkmark  &  \checkmark    \\ \hline
             180                    & \checkmark   & \checkmark    \\ \hline
             185                    & \checkmark   &  \checkmark   \\ \hline
             190                    & \checkmark   &  \checkmark    \\ \hline
             195                    & \checkmark   &  \checkmark    \\ \hline
             200                    & \checkmark   &  \checkmark   \\ \hline
		\end{tabular}
  \end{center}
\end{table} 

So far, we have only examined the presence or not of multipactor discharge for different electric field amplitudes. However, a relevant information to understand the magnitude of the discharge effects on the component is the electronic population growth factor. As it is well known \cite{Vaughan}, the multipactor phenomenon is characterised by an exponential growth of the electron population in the device that can be roughly described by the following expression: 

                                  \begin{equation} \label{ec_exponencial}
N_e(t/T)=N_0 \, e^{\sigma \frac{t}{T}} 
        \end{equation}
where $N_e$ is the number of electrons as a function of time, $N_0$ is the electron population at the initial instant, $T=1/f$ is the RF period $f$ being the RF frequency, and $\sigma$ is the growth factor which will depend, among other parameters, on the electric field amplitude $E_0$ and the SEY of the material the device walls are made of. The value of $\sigma$ can be obtained from fitting the data obtained in the numerical simulations and it is useful to make a rough estimate of how fast or slow the population growth is. Of course, the higher the value of $\sigma$ the faster the discharge will occur and the more noticeable its effects will be. However, it should be borne in mind that no matter how small the $\sigma$ is, the discharge will still occur if the field remains in the cavity for a sufficiently long time. 

\begin{figure*}
	\centering
	\includegraphics[scale=0.55]{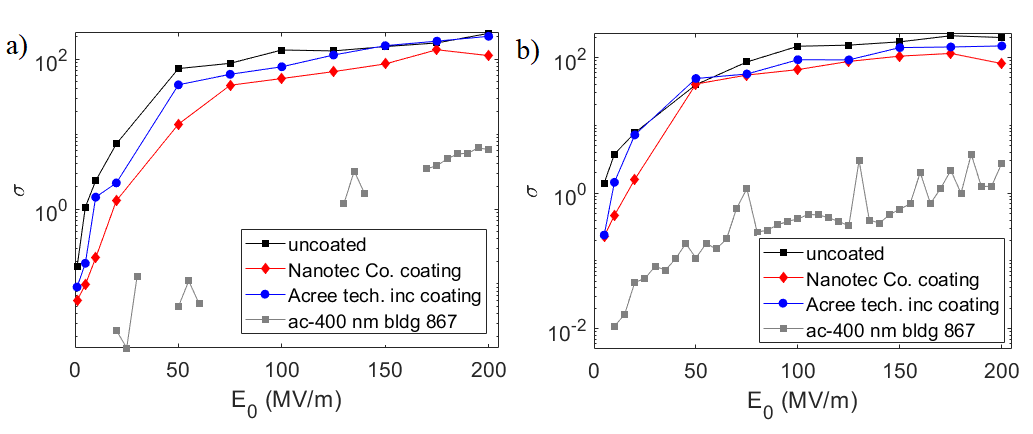}
	\caption{Comparisons of the growth factor $\sigma$ for the different dielectric material cases (dielectric uncoated, Nanotec Co. coating, Acree technologies inc. coating, ac-400 nm bldg 867 coating) for the different zones \textit{down} (a) and \textit{up} (b).  \label{sigma_comparativas}}
\end{figure*}

The $\sigma$ curves as a function of the electric field amplitude in the cavity are shown in Fig. \ref{sigma_comparativas} for each zone (\textit{up} and \textit{down}), and for the different types of dielectric materials. The results show that the general trend is that the higher the amplitude of the RF electric field, the faster the growth of the electron population. This phenomenon is observed in both the \textit{up} and \textit{down} zones. Regarding the comparison between the growth factors of the different dielectrics, it is observed that for the same value of $E_0$ the highest growth rate tends to be that of the dielectric without coating, followed in descending order by the Acree technologies inc. coating, the Nanotec Co. coating and, finally, by the ac-400 nm bldg 867 coating, which is the one with the lowest growth rate. This order is in line with the SEY curves as it would be expected, so that the lower the SEY values the lower the $\sigma$ values found.

\section{\label{an_model} Analytical approximated model for understanding the multipactor}

In this section we present an analytical model that, using a number of reasonable approximations, allows us to study the characteristics of the multipactor effect that occurs in the DAA structure. This model will be used below in the section \ref{phenomenology} to understand the results of the numerical simulations. First, we will focus on an electron departing from the radial wall of the cell. Remember that we define the radial walls of the structure as those where the normal vector to the wall coincides with the radial vector (see Fig. \ref{cell_zonas}). The RF electromagnetic field corresponding to the $TM_{02}$ mode present in the device has the following field components:

                          \begin{equation} \nonumber
\vec{E}_{RF}(\vec{r},t)=E_r(\vec{r},t) \hat{r}+ E_z(\vec{r},t) \hat{z} ;  \hspace{0.5cm} \vec{H}_{RF}(\vec{r},t)=H_{\phi}(\vec{r},t) \hat{\phi}
        \end{equation}
        
\begin{equation} \nonumber
\begin{split}
E_{r}=E_{0r}(r,z) \sin{(\omega t+\theta)}  \\ E_{z}=E_{0z}(r,z) \sin{(\omega t+\theta)} \\
H_{\phi}=H_{0}(r,z) \cos{(\omega t+\theta)}
\end{split}
\end{equation}
where $(r,\phi,z)$ are the radial, azimuthal and axial cylindrical coordinates, $(\hat{r},\hat{\phi},\hat{z})$ are the unit vectors in the radial, azimuthal and axial directions, respectively, $\omega = 2 \pi f$ is the angular frequency, and $\theta$ is the initial phase of the electromagnetic field.

An electron in the structure will be accelerated by the electromagnetic field inside it, so the force exerted on the particle is that given by the Lorentz force which, equated to Newton's Second Law in its non-relativistic version (we assume that the velocity of the electrons participating in the multipactor is much smaller than the value of the speed of light in vacuum), allows us to obtain the differential equations describing the motion of the electron. In our case, the differential equations for each component of the motion in cylindrical coordinates are described by:

\begin{equation} \label{ecs_dif}
\begin{split}
\frac{dv_r}{dt}=& -\frac{e}{m} E_{0r} \sin{(\omega t+\theta)} + \frac{e}{m} \mu_0 H_0 v_z \cos{(\omega t+\theta)}\\ \frac{dv_{\phi}}{dt}=& 0 \\
\frac{dv_z}{dt}=& - \frac{e}{m} \mu_0 H_0 v_r \cos{(\omega t+\theta)}  -\frac{e}{m} E_{0z} \sin{(\omega t+\theta)}
\end{split}
\end{equation}
where $v_r$, $v_{\phi}$, $v_z$, are the radial, azimuthal and axial components of the velocity vector, respectively, $e$ is the charge of the electron in absolute value, and $m$ is the mass of the electron. 

For the cases that we are going to consider we assume that the maximum displacement of the electron with respect to the starting position is small, or that the electron moves in a region where the electromagnetic field is approximately uniform. Thus, in the two previous cases, we can approximate the electromagnetic field that the electron sees in its trajectory by that of the initial position, and hence we will not take into account the inhomogeneity of the electromagnetic fields inside the DAA for the analysis of the electron's motion. Moreover, we can disregard in a first approximation in the third equation, corresponding to the time derivative of $v_z$, the first term which is proportional to the magnetic field $H_0$, since its contribution to the motion is weaker than the term with $E_{0z}$. By making this approximation, the resolution of the axial differential equation is notably simplified, since by integrating both sides of the equality with respect to time we can obtain the axial velocity as a function of time:

                          \begin{equation} \nonumber
\frac{dv_z}{dt} \approx -\frac{e}{m} E_{0z} \sin{(\omega t+\theta)} 
        \end{equation}

                                  \begin{equation} \nonumber
\int_{v_{0z}}^{v_z} dv_z = -\frac{e}{m} \int_0^t E_{0z} \sin{(\omega t+\theta)} dt
        \end{equation}

                          \begin{equation} \label{v_z}
v_z = v_{0z}+\frac{e}{m} \frac{E_{0z}}{\omega} \left[ \cos{(\omega t+ \theta)} - \cos{\theta} \right]
        \end{equation}

Substituting this expression into the equation for the radial coordinate we can again integrate on both sides of the equality with respect to time in order to obtain the radial velocity:

                          \begin{equation} \label{v_r}
                          \begin{split}
v_r= & v_{0r}+\frac{e}{m} \frac{E_{0r}}{\omega} \left[ \cos{(\omega t+ \theta)} - \cos{\theta}\right]  \\& +\frac{e}{m}\frac{\mu_0 H_0}{\omega} \left[ v_{0z} - \frac{e}{m} \frac{E_{0z}}{\omega}  \cos{\theta} \right] \left[ \sin{(\omega t+ \theta)} - \sin{\theta}\right] \\ &+ \left( \frac{e}{2\omega m} \right)^2 \mu_0 E_{0z} \left[ \sin{(2\omega t+2\theta)} +2\omega t -\sin{(2\theta)} \right]
\end{split}
        \end{equation}
        
The radial coordinate as a function of time is obtained with the integration of the radial velocity just obtained

                          \begin{equation} \label{ec_r}
                          \begin{split}
r= & r_0+\left[ v_{0r} -\frac{e}{m} \frac{E_{0r}}{\omega} \cos{\theta} - \frac{\xi'}{\omega}  \sin{\theta} -\frac{\xi}{4 \omega} \sin{(2\theta)} \right]t \\ &+ \frac{\xi}{4} t^2 +\frac{e}{m} \frac{E_{0r}}{\omega^2} \left[ \sin{(\omega t+ \theta)} - \sin{\theta} \right] \\ &- \frac{\xi'}{\omega^2} \left[ \cos{(\omega t+ \theta)} - \cos{\theta} \right]- \frac{\xi}{8 \omega^2}  \cos{(2\omega t+ 2\theta)} \\ & +\frac{\xi}{8 \omega^2} \cos{(2\theta)} 
\end{split}
        \end{equation}
    where we have entered the parameters $\xi= \left( \frac{e}{m}\right)^2 \frac{\mu_0 H_0 E_{0z}}{\omega}$ and $\xi'=\frac{e}{m} \mu_0 H_0 v_{0z} -\xi \cos{\theta}$. We are interested in the study of the single-surface multipactor with the radial wall, i.e. those electrons that initially leave the wall and, some time later, re-impact with the same wall at a position close to the starting position. Furthermore, we will assume that the time of flight of the electron between impacts, $\Delta t_i$, is much smaller than the period of the RF signal ($\Delta t_i/T<<1$). Under these conditions we can approximate the sine and cosine functions present in the equation (\ref{ec_r}) by Taylor series development around the initial phase $\theta$, 

                          \begin{equation} \nonumber
                          \begin{split}
\cos{(\omega t+ \theta)} \approx & \cos{\theta} - \omega t \sin{\theta} - \frac{(\omega t)^2}{2} \cos{\theta} \\
\sin{(\omega t+ \theta)} \approx &\sin{\theta} +\omega t \cos{\theta} - \frac{(\omega t)^2}{2} \sin{\theta} \\
\cos{(2(\omega t+ \theta))} \approx &\cos{(2\theta)} - 2\omega t \sin{(2\theta)} - 2(\omega t)^2\cos{(2\theta)}
\end{split}
        \end{equation}

                                  \begin{equation} \label{r_aprox}
r \approx r_0 + v_{0r}t + \frac{e}{2m} \left( \mu_0 H_0 v_{0z} \cos{\theta} - E_{0r} \sin{\theta}\right) t^2
        \end{equation}

The approximate equation for small radial path times (\ref{r_aprox}) corresponds to that of a uniformly accelerated rectilinear motion (UARM). For appropriate values of the initial phase of the field, the acceleration will have the opposite sign to the initial velocity so that the result of the motion will be that the electron will return to the exit surface shortly after emission. It is possible to estimate approximately the impact time at which the electron returns to the wall by imposing $r=r_0$ on the above equation obtaining,  

                                  \begin{equation} \label{t_aprox}
t_{i} \approx \frac{v_{0r}}{\frac{e}{2m} \left( E_{0r} \sin{\theta} - \mu_0 H_0 v_{0z} \cos{\theta}\right)}
        \end{equation}

        Analogously, we can estimate the angle of impact $\vartheta_i$ (in regard with the normal) using the expression 

                                  \begin{equation} \label{theta_aprox}
\vartheta_{i}  = \arccos{ \left( \frac{v_r(t_i)}{\sqrt{v_r^2(t_i)+v_z^2(t_i)}} \right)}
        \end{equation}
 where $v_r$ and $v_z$ are given by eqs. (\ref{v_r}) and (\ref{v_z}), respectively.

Similar equations can be derived for the case of an electron departing from the axial wall to study the single-surface multipactor in that case. Now, we will neglect in first approximation the term proportional to the magnetic field $H_0$ in the equation corresponding to the radial coordinate in (\ref{ecs_dif}), since it is weaker than the contribution proportional to $E_{0r}$. By making this simplification, we can integrate this equation directly and obtain the radial velocity as a function of time 

                          \begin{equation} \label{v_r_dos}
v_r = v_{0r}+\frac{e}{m} \frac{E_{0r}}{\omega} \left[ \cos{(\omega t+ \theta)} - \cos{\theta} \right]
        \end{equation}

Substituting this expression into the equation for the axial coordinate we can again integrate in both sides of the equality with respect to time for obtaining the axial velocity:

                          \begin{equation} \label{v_z_dos}
                          \begin{split}
v_z= & v_{0z}+\frac{e}{m} \frac{E_{0z}}{\omega} \left[ \cos{(\omega t+ \theta)} - \cos{\theta}\right] \\ &+ \frac{e}{m}\frac{\mu_0 H_0}{\omega} \left[  \frac{e}{m} \frac{E_{0r}}{\omega}  \cos{\theta} - v_{0r} \right] \left[ \sin{(\omega t+ \theta)} - \sin{\theta}\right]+  \\ & \left( \frac{e}{2\omega m} \right)^2 \mu_0 E_{0r} \left[ \sin{(2\omega t+2\theta)} +2\omega t -\sin{(2\theta)} \right]
\end{split}
        \end{equation}

The axial coordinate is calculated by integrating the axial velocity,

                          \begin{equation} \label{ec_z_dos}
                          \begin{split}
z=& z_0+\left[ v_{0z} -\frac{e}{m} \frac{E_{0z}}{\omega} \cos{\theta} - \frac{\eta'}{\omega}  \sin{\theta} -\frac{\eta}{4 \omega} \sin{(2\theta)} \right]t \\ & - \frac{\eta}{4} t^2 +\frac{e}{m} \frac{E_{0z}}{\omega^2} \left[ \sin{(\omega t+ \theta)} - \sin{\theta} \right] \\ &- \frac{\eta'}{\omega^2} \left[ \cos{(\omega t+ \theta)} - \cos{\theta} \right]+ \frac{\eta}{8 \omega^2}  \cos{(2\omega t+ 2\theta)}    \\ & -\frac{\eta}{8 \omega^2} \cos{(2\theta)}
\end{split}
        \end{equation}
 where we have introduced the parameters $\eta= \left( \frac{e}{m}\right)^2 \frac{\mu_0 H_0 E_{0r}}{\omega}$ and $\eta'=\eta\cos{\theta} -\frac{e}{m} \mu_0 H_0 v_{0r} $. If we assume that the electron time of flight between impacts is much smaller than the period of the RF signal, we can approximate the above equation by Taylor series to get

                                  \begin{equation} \label{z_aprox}
z \approx z_0 + v_{0z}t -\frac{e}{2m} \left( \mu_0 H_0 v_{0r} \cos{\theta} + E_{0z} \sin{\theta}\right) t^2 .
        \end{equation}

 It is possible to roughly estimate the impact time at which the electron returns to the wall by imposing $z=z_0$ on the above equation and clearing  

                                  \begin{equation} \label{t_aprox_z}
t_{i} \approx \frac{v_{0z}}{\frac{e}{2m} \left( E_{0z} \sin{\theta} + \mu_0 H_0 v_{0r} \cos{\theta}\right)}
        \end{equation}

        Similarly, we can estimate the angle of impact $\vartheta_i$ using the expression 

                                  \begin{equation} \label{theta_aprox_z}
\vartheta_{i}  = \arccos{ \left( \frac{ \pm v_z(t_i)}{\sqrt{v_r^2(t_i)+v_z^2(t_i)}} \right)}
        \end{equation}
        where $v_r$ and $v_z$ are given by eqs. (\ref{v_r_dos}) and (\ref{v_z_dos}), respectively. The negative sign is taken when the impact wall is the left sidewall and the positive sign for the right sidewall.

\section{\label{phenomenology} Multipactor phenomenology study}

In this section we will study in some detail the phenomenology of the multipactor discharge in the DAA cell for the case of the uncoated $MgTiO_3$ dielectric. The main characteristics of the multipactor are studied separately for the \textit{up} and \textit{down} zones of the DAA cell as a function of the RF electric field amplitude.

\subsection{\label{zone_down} Down zone}

\begin{figure*}
	\centering
	\includegraphics[scale=0.32]{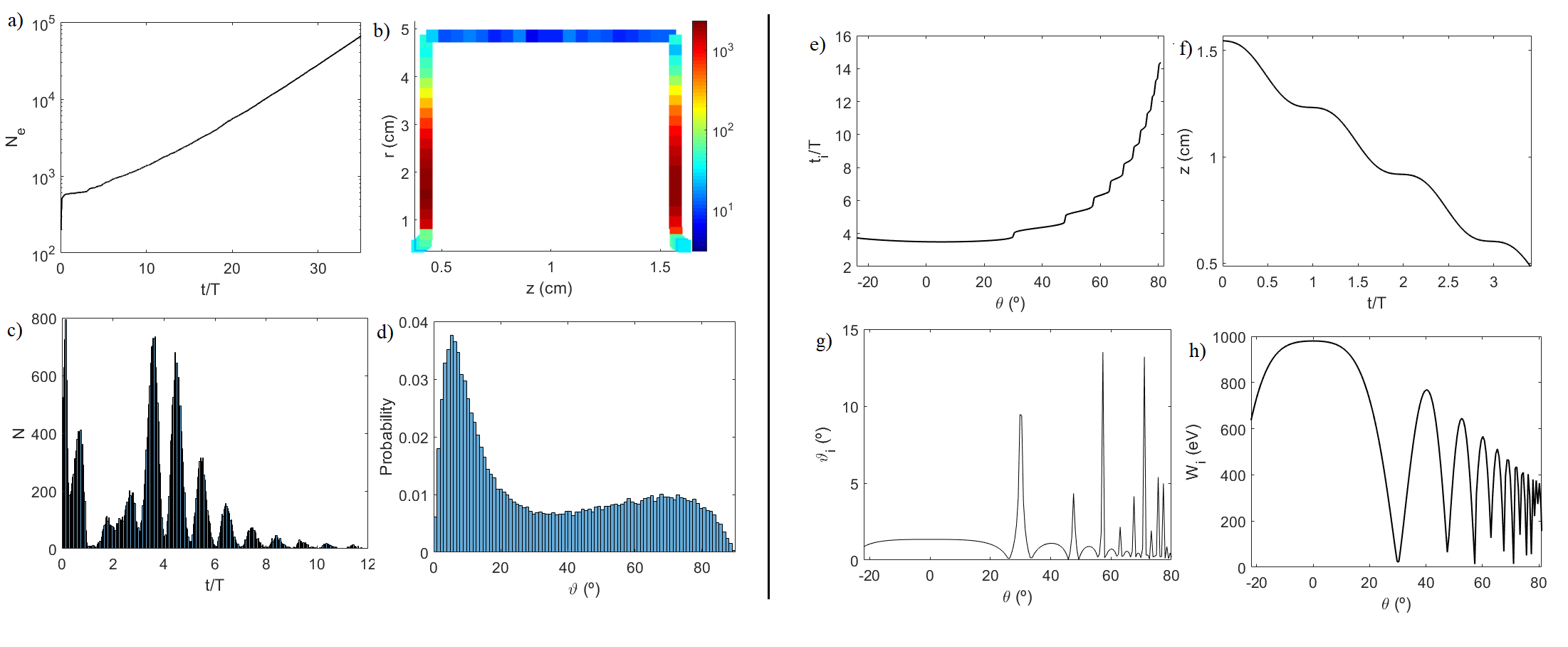}
	\caption{Left side: statistics of the multipactor simulation for $E_0 = 1$ MV/m (\textit{down} zone). a) Number of electrons in the structure as a function of time normalised to the period of the RF signal. b) Colour map with the number of electrons impacting at each position of the walls being able to generate two or more secondary electrons. c) Histogram with the time of flight of the electrons (generating two or more secondaries) between successive impacts normalised to the period of the RF signal. d) Histogram with the probability that an electron generating more than one secondary will impact at a certain angle with respect to the normal to the surface. Right side: results for $E_0 = 1$ MV/m (\textit{down} zone) from the theoretical model for an electron departing from the right lateral wall and impacting with the left sidewall. e) Electron impact time normalized to the RF period as a function of initial field phase. f) Radial trajectory for an electron with initial field phase $\theta = 10^{\circ}$. g) Impact angle of the electron with respect to the surface normal as a function of the initial field phase. h) Impact kinetic energy of the electron as a function of the initial field phase.  \label{1MVm-down}}
\end{figure*}

We start by studying the smallest value of electric field for which discharge was detected in the \textit{down} zone, i.e., $E_0 = 1$ MV/m, whose results in terms of electron statistics extracted from the multipactor simulations are shown in Fig. \ref{1MVm-down} (a-d). In Fig. \ref{1MVm-down} a) it is shown the electron population as a function of the time normalized to the RF period. Fig. \ref{1MVm-down} b) shows the spatial distribution of electrons colliding with the dielectric walls of the structure, considering only electrons that are capable of generating two or more true secondaries. This information is useful to determine the region where the multipactor discharge originates. In this case we see that the main areas where secondary electron emission occurs are both sidewalls, with a maximum contribution in the central region of the lower halves that decreases as the radial coordinate increases as we approach the radial wall. The electron flight time between successive impacts is shown in the histogram of Fig. \ref{1MVm-down} c). In the simulations, it is observed that the resonant electron trajectories are both single-surface and double-surface. In the single-surface trajectories the electron leaves the exit surface and re-impacts in an area close to it in a time much shorter than the period of the RF signal. This type of trajectories would correspond to the peak in the impact time distribution that appears around $t_i/T=0.15$ in Fig. \ref{1MVm-down} c). With respect to the double-surface trajectories, these occur between the sidewalls at times that coincide with the typical half-integer orders described in the well-known classical multipactor theory \cite{Vaughan}. Indeed in Fig. \ref{1MVm-down} c) peaks in the histogram are observed for times between impacts $t_i/T$ of $2.5, 3.5, 4.5, 5.5$, etc. The angles of incidence of the electrons with respect to the normal to the impact surface are plotted in Fig. \ref{1MVm-down} d). The distribution of electron impact angles shows its main maximum at an angle of about $\vartheta = 5^{\circ}$, although the tail of the curve extends to larger angles, with a small fraction of electrons colliding even at angles greater than $\vartheta = 80^{\circ}$. 

The characteristics of the multipactor for $E_0 = 1$ MV/m can also be studied with the help of the analytical model presented in section \ref{an_model} and compared with the results found in the simulations. We will examine the particular case of an electron exiting the right sidewall in the region where the simulations predict higher population growth and look for double-surface resonant trajectories, i.e., that the electron ends up impacting the left sidewall. The analytical model predicts electron flight times between the sidewalls as a function of the initial phase of the field shown in Fig. \ref{1MVm-down} e). It is observed that there is a wide range of initial phases, from  $\theta = -25^{\circ}$ to $\theta = 30^{\circ}$, in which the time of flight remains fairly stable around the mean value $t_i/T = 3.5$. This phase range agrees with the local maximum of the distribution found in the simulations (see Fig. \ref{1MVm-down} c)). Returning to Fig. \ref{1MVm-down} e), for initial phases in the range $\theta = 31^{\circ}-47^{\circ}$ a new plateau appears in the curve with impact times around the mean value $t_i/T=4.5$, which corresponds to another of the local maxima of the distribution shown in Fig. \ref{1MVm-down} c). Notice that in Fig. \ref{1MVm-down} e) as the initial phase increases steps in the impact times around values of $t_i/T=5.5, 6.5, 7.5,$ etc. keep appearing, which of course coincide with the typical orders of the classical double-surface multipactor theory that were also found in the simulations. Furthermore, in Fig. \ref{1MVm-down} e) one has that the phase range corresponding to each multipactor order (each of the steps) is narrower, which is related to fewer electrons being available to contribute to the discharge and agrees with that in Fig. \ref{1MVm-down} c) the amplitude of the peaks of the distribution decays as the electron flight time increases. Also the impact angles predicted by the analytical model (see Fig. \ref{1MVm-down} g)), which are mostly below $\vartheta= 5^{\circ}$, are compatible with the distribution shown in Fig. \ref{1MVm-down} d) which has a large fraction of collisions with angles close to zero. Finally mention that the electron impact energies (see Fig. \ref{1MVm-down} h)) are above the $W_1$ value of the material at normal incidence, and hence
electrons will generate secondaries for any initial electron phase.

\begin{figure*}
	\centering
	\includegraphics[scale=0.32]{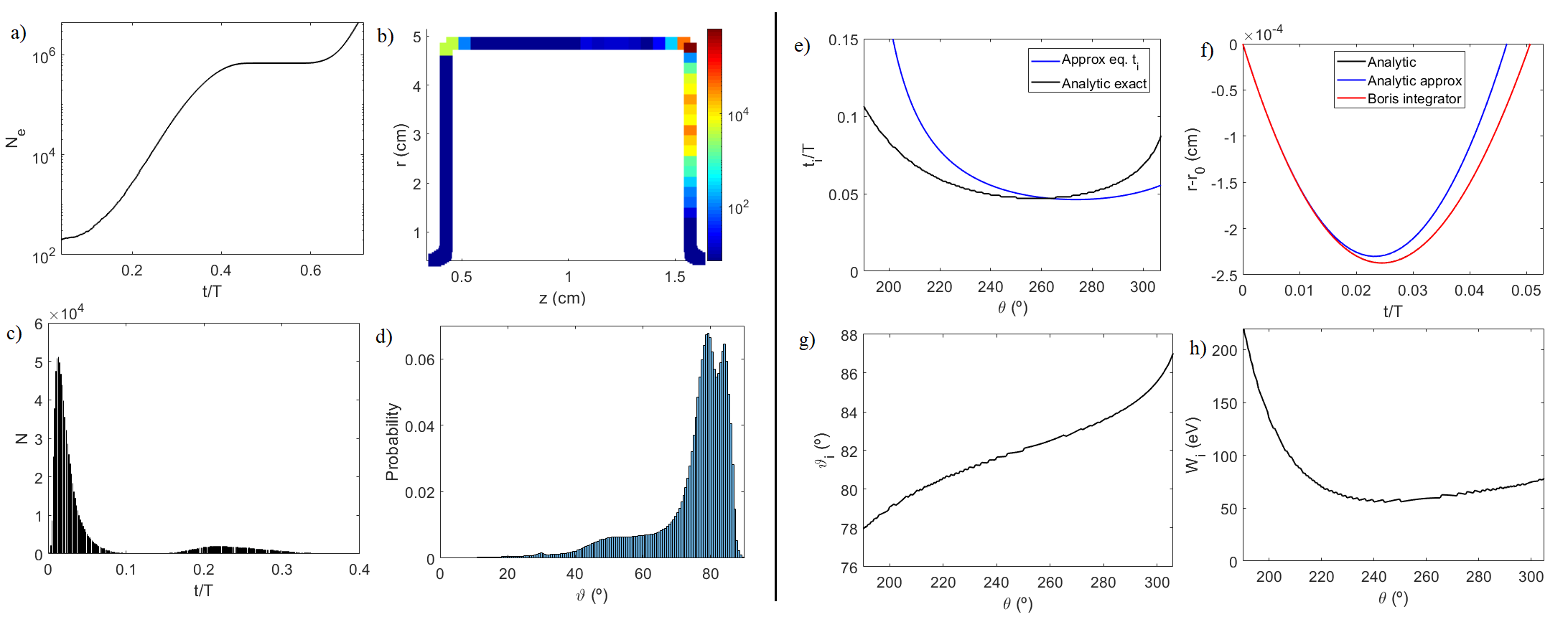}
	\caption{Left side: statistics of the multipactor simulation for $E_0 = 20$ MV/m (\textit{down} zone), same captions as in Fig. \ref{1MVm-down}. Right side: results for $E_0 = 20$ MV/m (\textit{down} zone) from the theoretical model for an electron departing from the lateral wall and impacting again with the same surface. e) Electron impact time normalized to the RF period as a function of initial field phase, including the results provided by the approximate eq. approximate (\ref{t_aprox}) and by the trajectory equation (\ref{ec_r}). f) Radial trajectory for an electron with initial field phase $\theta = 280^{\circ}$, including the results of the analytical eq. (\ref{ec_r}), of the approximate eq. (\ref{r_aprox}), and of the differential eq. of the motion integrated with the Boris method. Same captions for g) and h) as in Fig. \ref{1MVm-down}. \label{20MVm-down}}
\end{figure*}

The next case to be examined in detail is that of $E_0 = 20$ MV/m. The data from the simulations are shown in the  Fig. \ref{20MVm-down} (a-d). In Fig. \ref{20MVm-down} b) it is seen that there are three clearly predominant regions: the region near the left corner of the dielectric (both in the radial and axial part of the wall) the region analogous to the previous one corresponding to the right corner and, with a smaller proportion of impacts than the two previous ones although more extensive in surface, the upper half (approximately) of the side walls. It should be mentioned that although in Fig \ref{20MVm-down} b) impacts are only observed on the right side, they are also expected to occur symmetrically on the left side of the side wall. The reason why they do not appear in our simulations is simply because the simulation time is shorter than the period of the RF signal, and therefore too short to observe the effect of electron displacement from one sidewall to the opposite sidewall. The simulation time is shortened due to the rapid growth of the electron population, which causes the number of electrons to be high after a short time interval, thus making the simulation increasingly slower due to the number of particles to be considered. The electron flight time between successive impacts is shown in the histogram of Fig. \ref{20MVm-down} c). The vast majority of electrons have impact times shorter than $t_i/T=0.1$, with the maximum of the distribution centred approximately around $t_i/T=0.05$. This type of multipactor is non-resonant because there is no synchronization between the RF electric field and the electron trajectory. In the classical multipactor theory, the electron must have a time of flight between successive impacts with the component walls of a odd (even) number of half-periods of the RF signal to guarantee the periodicity of the electron trajectory and the development of the double (single) surface multipactor. In the single-surface discharge found here, the multipactor order is much smaller than the minimum value predicted by theory (two RF half-periods), so the resonance between the electron trajectory and the RF electric field does not occur. Despite this, as the electron impacts with sufficient energy to release secondary electrons, the electron population will increase and the discharge will eventually occur. Since the time between successive impacts is much shorter than the typical times predicted by classical multipactor theory, there will be many collisions in each RF period and the population growth will be much faster than for classical multipactor orders. Because of this we will call this type of discharge as non-resonant ultra-fast multipactor. The angles of incidence of the electrons with respect to the normal to the impact surface are plotted in Fig. \ref{20MVm-down} d), finding a prevalence of oblique angles $\vartheta> 40^{\circ}$ with the most likely value slightly exceeding $80^{\circ}$. 

The characteristics of the multipactor found in the simulations for $E_0 = 20$ MV/m can be justified on the basis of the approximate equations of motion of the electrons in the structure that we have derived in section \ref{an_model}. Let us focus on the trajectories of the electrons leaving the radial wall in the region near the right-hand corner, which corresponds to one of the hot regions for the multipactor. Using the equations of motion (\ref{v_z})-(\ref{theta_aprox}) we can obtain the trajectory and the main parameters of the electron dynamics inside the DAA cell, as it is shown in Fig. \ref{20MVm-down} (e-h). In Fig \ref{20MVm-down} e) the time taken for the electron to hit the walls is plotted normalised to the RF period and expressed as a function of the initial phase of the field. The results have been plotted using for this calculation the equation of the radial trajectory (\ref{ec_r}), referred to in the legend as exact analytical, and the approximation of this equation for small times $t/T<<1$ (\ref{t_aprox}). Eq. (\ref{ec_r}) shows that for a wide range of initial phases in the approximate range $180^{\circ}-305^{\circ}$ the electron hits at times between $t/T=0.05$ and $t/T=0.1$, i.e., much smaller than the RF period. On the other hand, we note that eq. (\ref{t_aprox}) gives a reasonable but not very accurate first approximation of the collision time. In Fig. \ref{20MVm-down} f) the radial trajectory is plotted as a function of normalised time for an electron with initial phase $\theta = 280^{\circ}$. The values of the analytical solution of eq. (\ref{ec_r}) and eq. (\ref{r_aprox}) which is an approximation of the previous one for short times, and the solution of the numerical integration of the differential equations of motion (\ref{ecs_dif}) using the Boris method \cite{Ripperda_2018} have been plotted. The first thing to note from this plot is that the electron describes a typical single-surface multipactor trajectory, since shortly after being emitted it re-impacts with the surface from which it was emitted. The short time of flight implies that the electron separates little from the exit surface, a fact that is in agreement with the approximation made above of taking a uniform value of the fields to solve analytically the differential equations of motion of the electron. On the other hand, eq. (\ref{ec_r}) and Boris method agree perfectly with each other, a fact that allows us to justify and validate the simplification made in the third equation of the system (\ref{ecs_dif}) of disregarding the term proportional to the magnetic field $H_0$. The analytical approximation (\ref{r_aprox}) gives reasonable values for the electron trajectory, although the fact that its validity is restricted to very small times means that as time goes by the approximate trajectory differs from the full equation (\ref{ec_r}). In Fig. \ref{20MVm-down} g) we have the impact angle with respect to the normal to the surface as a function of the initial phase of the electromagnetic field. The impact angles in the studied phase interval are in the range $78^{\circ}-87^{\circ}$, i.e., they are very oblique collisions. Finally, in Fig. \ref{20MVm-down} h) the electron impact kinetic energy is plotted as a function of the initial phase of the field. Over the whole range of phases analysed, the impact kinetic energy greatly exceeds the value of $W_1 = 28.6$ eV of the material at normal incidence and, therefore, will also exceed $W_1$ at any angle above this. This indicates that electrons leaving the wall in the above phase range will be within a short time return to impact with the wall and will do so by releasing new secondary electrons. In addition, the oblique incidence increases the number of emitted electrons because the SEY always increases as the angle of incidence increases. In conclusion, during the phase interval of the RF field shown in Fig. \ref{20MVm-down} e) the necessary conditions will exist for single-surface multipactor to occur on the radial wall in the area near the right-hand corner. The characteristics of this multipactor regime (type of trajectories, time between impacts, impact angle) are in full agreement with those found in the simulations shown in Fig. \ref{20MVm-down} (a-d). During each period of the RF signal there will be a certain time interval (or equivalently initial field phases) during which the electromagnetic field will not be favourable for the electron to return to the exit surface and it will end up being pushed towards other regions of the dielectric cell. In many cases, the electron will find another region on another surface of the dielectric where conditions will be appropriate for new single-surface multipactor trajectories to be established. In the example that we are dealing with, the multipactor process can be summarized by the following steps. Initially the electrons in the region near the right-hand corner of the radial wall are in the phase range of the field appropriate for generating single-surface multipactor. Some time later the field polarity change pushes them into the lateral region of the right-hand wall also close to the corner. Here a new single-surface multipactor is established  until the orientation of the electromagnetic field becomes unfavourable again and pushes them out of the wall, so that many of these electrons will move towards the first surface (radial wall near the right-hand corner) from which they originally started, thus completing a periodically repeating cycle with a net increase of the electron population. In addition to the single-surface multipactor moving between the radial and side walls of the corners described above, there is also a double-surface multipactor between the top of the side walls, but with a smaller contribution to electron population growth than the single-surface one. 

The statistics of the multipacting electrons were also examined for other RF electric field amplitudes in the range from $1$ to $200$ MV/m. In addition to the $1$ MV/m and $20$ MV/m cases described above, results for $E_0=[5,10,50,75,100,150,200]$ MV/m were also analyzed. For the sake of brevity we will summarize the main features that were found for the multipactor behavior in the \textit{down} zone:

\begin{itemize}
    \item The zone where the discharge originates at low electric field amplitudes ($E_0 = 1$ MV/m) is concentrated in the lower half of the sidewalls. As $E_0$ increases, this zone shifts towards the upper half of the same sidewalls and appears in the region near the upper corners (both radial and lateral walls). This zone transition occurs in the interval $1-10$ MV/m. For $E_0>10$ MV/m the multipactor regions are hardly changed.

    \item The resonant electron trajectories are of both the single-surface and double-surface types. However, as $E_0$ increases, the contribution to electron growth of the single-surface multipactor increases relative to the double surface type.
    
    \item The time of flight of electrons between successive impacts tends to decrease with increasing $E_0$. For $E_0 = 10$ MV/m, where single-surface multipactor is dominant, the peak in the time of flight distribution is around $t_i/T= 0.02$. For $E_0=20$ MV/m the maximum is at $t_i/T=0.05$ and for $E_0 =200$ MV/m the peak shifts to $t_i/T= 0.002$. 

    \item At low field amplitudes ($E_0 = 1$ MV/m) most electrons impact at small angles to the normal. However, as $E_0$ increases, an increase in the collision angle is observed. For fields with $E_0 \geq 10$ MV/m the maximum of the impact angle distribution is around $\vartheta = 70^{\circ}$.  
\end{itemize}

\subsection{\label{zone_up} Up zone}

\begin{figure*}
	\centering
	\includegraphics[scale=0.32]{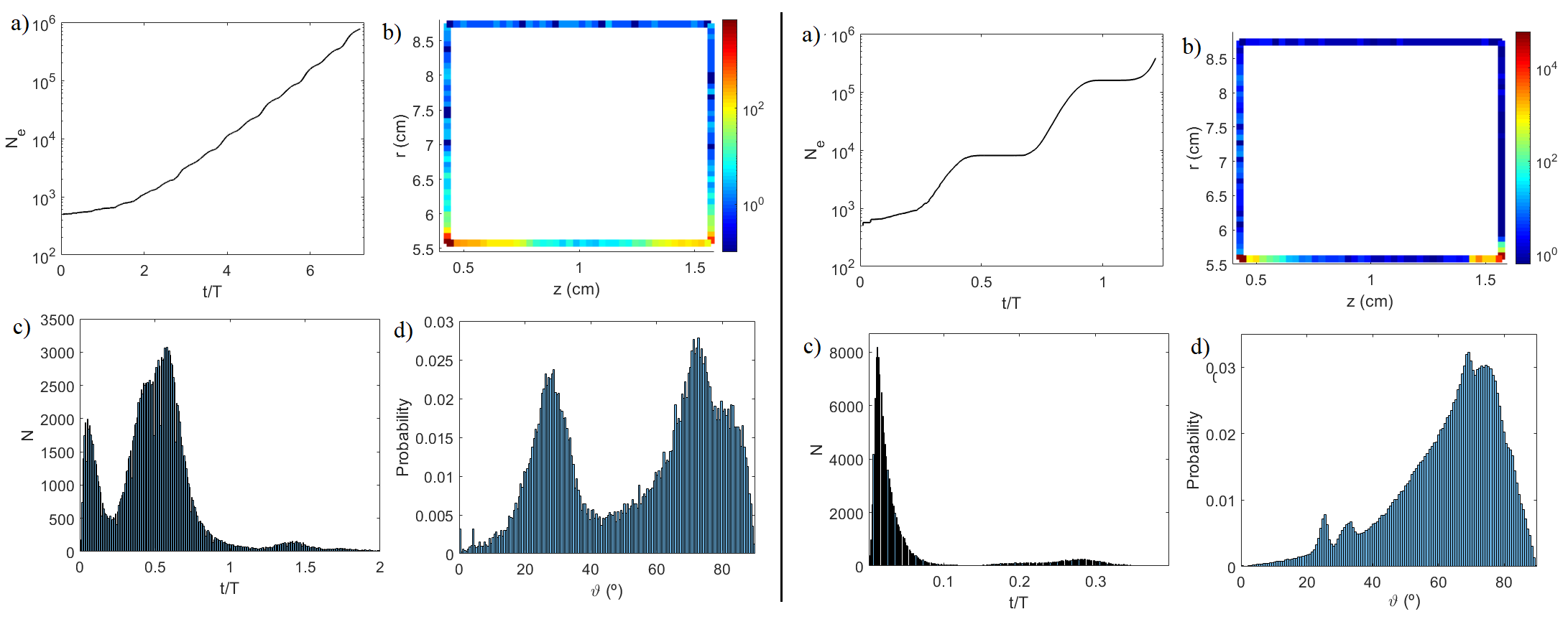}
	\caption{Left (right) side: statistics of the multipactor simulation for $E_0 = 5$ MV/m ($E_0 = 20$ MV/m) in the \textit{up} zone. Same captions as in Fig. \ref{1MVm-down}.  \label{5-20MVm-up}}
\end{figure*}

The smallest value of electric field for which discharge was detected in the \textit{up} zone is $E_0 = 5$ MV/m, and the electron statistics from the multipactor simulations are shown in the left side of the Fig. \ref{5-20MVm-up}. The regions where most of the impacts that are able to generate new secondaries occur correspond to the regions near the lower left and lower right corners. It is worth mentioning that although secondary electron generation can also occur at other positions of both the sidewall and the top wall, the probability is lower and the resulting electron trajectories are usually not resonant for the multipactor. As for the time of flight between impacts in the resonant trajectories, the distribution has its main maximum around $t/T=0.5$, another local maximum around $t/T=0.05$, and decays rapidly for $t/T>1$. These resonant trajectories are largely of the single-surface type, in which the electron eventually re-impacts the exit surface, at a position relatively close to the initial one. With regard to the angle of impact, the main peak of the distribution is around $\vartheta = 70^{\circ}$, plus a secondary maximum appears around $\vartheta = 30^{\circ}$. Therefore, there is a predominance of oblique angles of incidence. All these results are in agreement with the theoretical model in section \ref{an_model}, but a more detailed comparison is not presented here for the sake of brevity. 

The next scenario that we study corresponds to $E_0 = 20$ MV/m. The simulation statistics are detailed in the right side of the Fig. \ref{5-20MVm-up}. The regions where the discharge occurs remain very similar to those found for $E_0 = 5$ MV/m, i.e. mainly the areas close to the lower left and right corners. The time of flight between impacts has been reduced with respect to the previous case, now the maximum of the distribution is around $t/T=0.012$ and no trajectories with times longer than $t/T=0.4$ are found. Variations are also observed in the impact angles, which become more oblique, although the main peak of the distribution is still at around $\vartheta= 70^{\circ}$, the secondary maximum, which previously appeared at around $\vartheta= 30^{\circ}$, has been significantly reduced to the detriment of the main peak. 

The statistics of the multipacting electrons in the \textit{up} zone were also examined for $E_0=[10,50,75,100,150,200]$ MV/m. The main characteristics that were found for the multipactor behavior in  this zone can be summarized as follows: 

\begin{itemize}
    \item The areas where the discharge is generated are primarily those near the lower left and right corners. These regions remain fairly stable in the range $E_0 =5-200$ MV/m.
        \item  Single-surface resonant trajectories are highly predominant. 
    \item The time of flight of electrons between successive impacts tends to decrease with increasing $E_0$. For $E_0=5$ MV/m, the maximum of the time of flight distribution is around $t_i/=0.5$, for $20$ MV/m is $t_i/T=0.012$, for $100$ MV/m is $t_i/T=0.002$, and for $200$ MV/m is $t_i/T=0.001$.
    \item The impact angle of electrons with surfaces becomes more oblique as $E_0$ increases. Even for low values of $E_0$ typical collision angles depart from normal incidence. For $E_0 = 200$ MV/m  almost all electrons collide at angles greater than $\vartheta = 80^{\circ}$.
\end{itemize}

\section{\label{mod_cell} Modification of the DAA cell for multipactor mitigation}

\begin{figure}[H]
\centering
\includegraphics[scale=0.4]{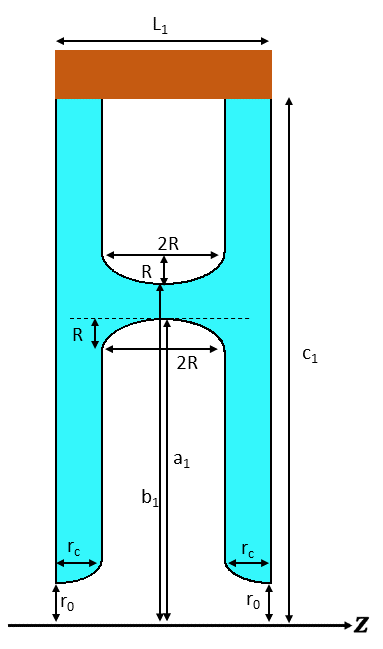}
\caption{Schematic of the modified DAA cell. \label{DAA_scheme_modified}}
\end{figure}

As shown before for many of the dielectrics considered the appearance of the multipactor was significant for a wide range of electric field amplitudes. In this section we propose to modify the geometry of the DAA cell in order to reduce as much as possible the occurrence of this undesired phenomenon. According to the simulations, in many cases, the region of the DAA cell where the discharge develops the most is around the corners of the structure. Therefore, it seems reasonable to assume that if the corners of the dielectric are somehow smoothed, this could have a negative effect on the multipactor resonant trajectories, which in certain cases would hinder the development of the discharge. Based on this hypothesis, the DAA cell was redesigned to change the straight sections of the radial dielectric walls to circular sections, as shown in Fig. \ref{DAA_scheme_modified}. The radius of curvature $R$ has to satisfy that $R=\frac{L_1-2r_c}{2}$. For the new design, the dimensions of the original DAA cell were maintained except for $b_1$, whose variation was necessary to adjust the resonance frequency to the desired value. In our case, with $b_1 = 5.465$ cm we obtain the mode frequency $f= 2999.23$ MHz. 

Below, the results of the multipactor simulations for the modified cell with the different dielectric materials are shown below, showing also the comparison with the results of the original cell, in order to be able to evaluate the possible benefits of the new design.

\subsection{Dielectric $MgTiO_3$ without coating}

\begin{figure*}
	\centering
	\includegraphics[scale=0.49]{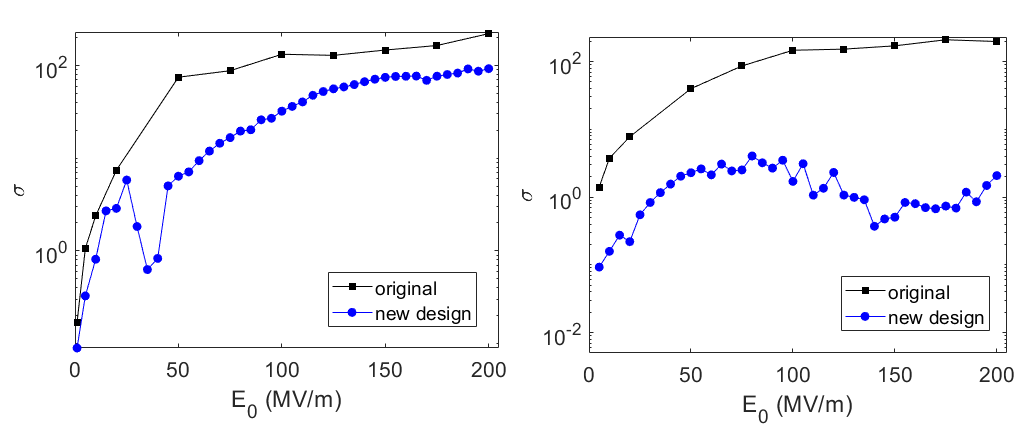}
	\caption{Growth factor comparisons for uncoated $MgTiO_3$ in the original cell and in the new design, for the \textit{down} (left) and \textit{up} (right) zones.  \label{sigma_uncoated_nuevageom_vs_old}}
\end{figure*}

Numerical simulations for the modified DAA structure with the uncoated $MgTiO_3$ dielectric show that the multipactor appears for the same range of electric field amplitude as in the original cell design, i.e., the multipactor threshold is $E_0 = 1$ MV/m in the \textit{down} zone and $5$ MV/m in the \textit{up} zone. Hence, no improvement is found with the new geometry in terms of the multipactor presence in the device. Similarly as it was done for the original design, it is interesting to obtain from the simulations the $\sigma$ factor which gives an idea of how fast the electronic population grows. In Fig. \ref{sigma_uncoated_nuevageom_vs_old} the values of $\sigma$ for the \textit{down} and \textit{up} zones are shown and compared with the results of the original geometry. It is found that in both zones the new design has a lower growth factor than the original design so that, although the new geometry is not able to inhibit the discharge, it is able to partially hinder its occurrence. 

\subsection{Dielectric $MgTiO_3$ with Nanotec Co. coating}

The results of the multipactor simulations for the modified DAA structure with the dielectric $MgTiO_3$ with Nanotec Co. coating show that the RF electric field multipactor threshold remains the same as in the original cell design with this coating (i.e., $1$ MV/m in the \textit{down} zone and $5$ MV/m in the \textit{up} zone). However, in the \textit{down} area, a wide window of multipactor-free RF electric field amplitudes appears in the range $35-125$ MV/m. On the other hand, in the \textit{up} zone we do not get any improvement with the geometry modification and the discharge appears at exactly the same values of $E_0$ as for the original design.

\begin{figure*}
	\centering
	\includegraphics[scale=0.49]{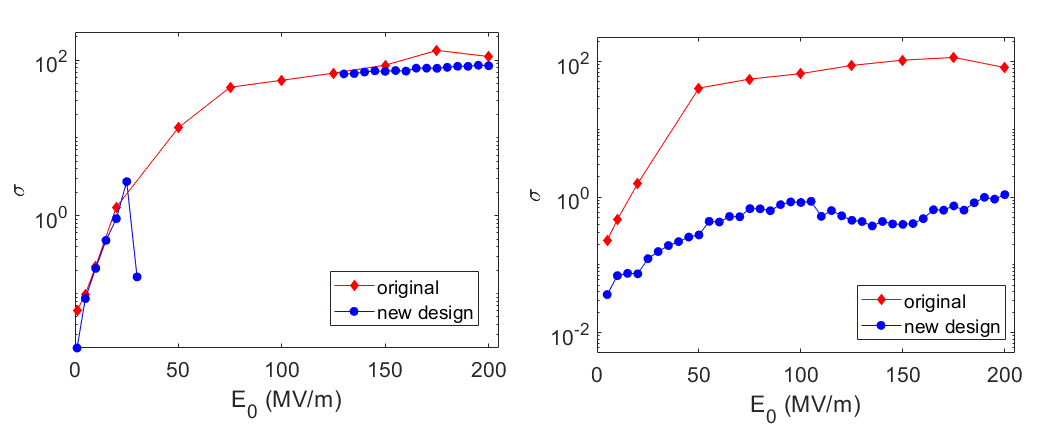}
	\caption{Comparisons of the growth factor $\sigma$ for the Nanotec Co. coating in the original cell and in the new design, for the \textit{down} (left) and \textit{up} (right) zones.  \label{sigma_japonesa_nuevageom_vs_old.png}}
\end{figure*}

The comparison of the growth factor $\sigma$ between the original and the modified design is shown in Fig. \ref{sigma_japonesa_nuevageom_vs_old.png}. In the \textit{down} zone the $\sigma$ values, where discharge appears, are similar in both designs, although they tend to be slightly lower in the new prototype. With respect to the \textit{up} zone, a significant decrease in $\sigma$ is observed in the case of the new design. In conclusion, the modification of the geometry does represent an appreciable improvement for the Nanotec Co. coating, since it is capable of generating $E_0$ windows free of multipactor in the \textit{down} zone and of reducing the discharge growth factor in the \textit{up} zone, although in the latter case it is not capable of preventing the appearance of the discharge.

\subsection{Dielectric $MgTiO_3$ with ac-400 nm bldg 867 coating}

The simulations for the modified DAA structure with the dielectric $MgTiO_3$ with the ac-400 nm coating show that in the \textit{down} region multipactor only appears for $E_0 = 25$ MV/m, which is a significant improvement in this region compared to the original cell, where multipactor appeared from $20$ MV/m, with multipactor-free windows in the $35-45$ MV/m, $65-125$ MV/m, and $145-165$ MV/m ranges. However, in the \textit{up} zone the results are identical to those shown for the original cell, and no improvement in discharge suppression is found.

\begin{figure*}
	\centering
	\includegraphics[scale=0.49]{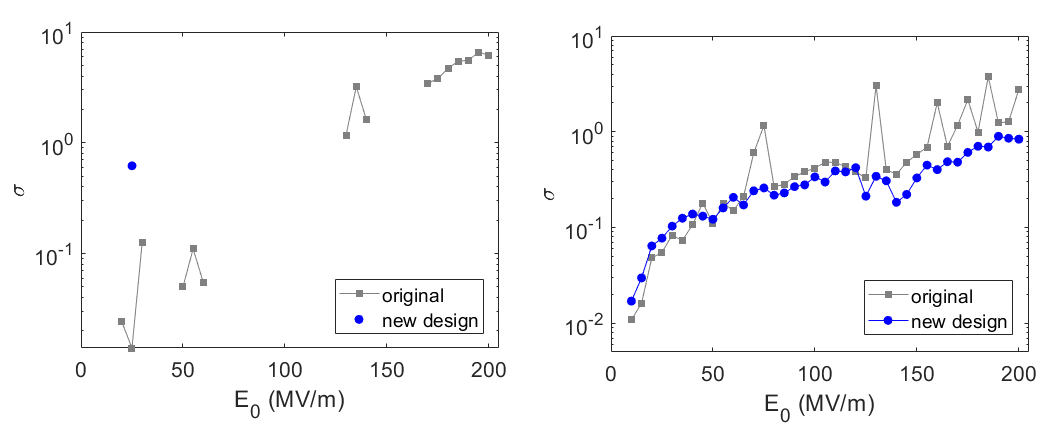}
	\caption{Comparisons of the growth factor $\sigma$ for ac-400 nm bldg 867 coverage in the original cell and in the new design, for the \textit{down} (left) and \textit{up} (right) zones.  \label{sigma_ac40nnbldg867_nuevageom_vs_old}}
\end{figure*}

The comparison of the growth factor $\sigma$ between the two geometries of the DAA cell can be seen in Fig. \ref{sigma_ac40nnbldg867_nuevageom_vs_old}. In the \textit{down} zone for the new geometry only discharge appears for a value of $E_0$, where the growth factor is higher in the new geometry with respect to the original. In the \textit{up} zone the $\sigma$ curves are quite similar in both geometries. Similar to the case of the Nanotec Co. coating, the improvement of the new geometry with respect to the original design is observed, since in the \textit{down} zone it is able to almost completely inhibit the multipactor discharge for any value of electric field amplitude (discharge only appears for $25$ MV/m) and in the \textit{up} zone relatively low levels of $\sigma$ are maintained.

\section{\label{conclusions} Conclusions}

In this work, the risk of suffering a multipactor discharge in an S-band dielectric-assisted accelerator (DAA) structure for a compact low-energy linear particle accelerator has been studied by numerical simulations performed with an in-house developed code. The results of our code have been compared with the commercial software CST Particle Studio, finding good agreement between them and thus verifying our algorithm. Multipactor analysis was performed for the dielectric material as fabricated, and for various coatings that tend to reduce secondary electron emission. The multipactor simulations explored a range of RF electric field amplitudes from $0.01$ to $200$ MV/m and the presence of discharge was evidenced for the uncoated dielectric in the range of $1$ to $200$ MV/m in the \textit{down} zone, and $5$ to $200$ MV/m in the \textit{up} zone. The results for the coatings from the Nanotec Co. and Acree technologies inc. were the same as those for the uncoated dielectric. However, for the ac-400 nm bldg coating it was found that the multipactor threshold, in the \textit{down} zone, appears at $E_0 = 20$ MV/m and there are multipactor-free windows in the ranges $35-45$ MV/m, $65$-$125$ MV/m, and $145$-$165$ MV/m. On the other hand, in the \textit{up} zone this coating raises the threshold to $10$ MV/m but no multipactor-free windows appear. The phenomenology of the multipactor in the cavity (types of electron resonant trajectories, zones of the cell where the discharge appears, time of flight between successive electron impacts with the walls, impact angle with respect to the surface) was analyzed in detail for the uncoated case with the help of a theoretical model developed from analytically solving the differential equation of motion of electrons using some approximations. The theoretical model was found to show good agreement with the simulation results. It is worth mentioning that for many amplitudes of the RF electric field in the cavity a new non-resonant and ultra-fast multipactor has been found that has not been still described by classical multipactor theory. On the other hand, since the presence of the multipactor in the DAA cell has been shown to be of concern, we proposed a modification of the design geometry in order to reduce as much as possible the occurrence of this unwanted phenomenon. For this new design and with the Nanotec Co. coating, it was found that the multipactor threshold of the RF electric field in both zones remains the same as in the original cell design with this coating. However, in the \textit{down} zone, a wide window of multipactor-free RF electric field amplitudes appears in the range $35$-$125$ MV/m. Although no multipactor-free windows appear in the \textit{up} zone, a decrease of the electron population growth factor is observed. With respect to simulations of the new geometry with the ac-400 nm bldg 867 coating, in the \textit{down} zone the multipactor appears only around $E_0 = 25$ MV/m and is completely suppressed for other electric field amplitudes. In the \textit{up} zone the multipactor still appears for the same values of $E_0$.

Therefore, it is shown that geometry modification helps to prevent the occurrence of discharge, which needs to be used in combination with the application of low SEY coatings to the dielectric. Thus, the RF power handling of the accelerating structure can be increased by these techniques and a correct multipactor risk assessment on the DAA structure.

\bibliography{references}

\end{document}